\documentstyle[12pt,epsf]{article}

\begin{document}
\begin{titlepage}
\begin{center}

{\Large \bf A Non-perturbative Solution of \\ the Zero-Dimensional 
$\lambda\varphi^{4}$ 
Field Theory}\\
\vspace{.3in}
{\large\em A.P.C. Malbouisson
\footnotemark[1], R. Portugal \footnotemark[2], and N.F. Svaiter
\footnotemark[3]}
\\
Centro Brasileiro de Pesquisas F\'{\i}sicas-CBPF\\ Rua Dr.Xavier
 Sigaud 150, Rio de Janeiro, RJ 22290-180 Brazil\\ 

\end{center}

\begin{abstract}
We have done a study of the 
zero-dimensional $\lambda\varphi^{4}$ model.
Firstly, we exhibit the partition 
function as a simple exact expression in terms of the 
Macdonald's function for $Re(\lambda)>0$. Secondly, an analytic 
continuation of the partition function for $Re(\lambda)<0$
is performed, and we
obtain an expression defined in the  
complex coupling constant plane $\lambda$, for $|arg\,\lambda|<\pi$.
Consequently,  the 
partition function understood as an analytic continuation is defined for all values of 
$\lambda$, except for a branch cut along the real 
negative $\lambda$ axis. We also evaluate the partition function
on perturbative grounds, using the Borel summation technique and
we found that in the common domain of validity for $Re(\lambda)>0$, 
it coincides precisely with the exact expression. 
\end{abstract}

\


  
\footnotetext[1]{e-mail:adolfo@lafex.cbpf.br}
\footnotetext[2]{e-mail:portugal@cbpf.br} \footnotetext[3]{e-mail:nfuxsvai@lafex.cbpf.br}

\end{titlepage}
\newpage
\baselineskip .37in
\section {Introduction}

It is largely accepted that 
some insight on the behavior of Green's functions in Field theories and Statistical systems may be obtained by the analysis of zero-dimensional models \cite{novas1} \cite{novas2}. In particular it is expected that an exact understanding of some aspects of zero-dimensional field
theories could leave trails about the non-perturbative behaviour of these theories
in higher dimensions.
 In this sense, Bender et al. \cite {Bender1} for instance, have proposed an analytical 
approach to study non-perturbativelly quantum field theories, which 
requires to solve the corresponding zero-dimensional model.
In the context of perturbative field theory, for the study 
of the summability of series giving evaluations of physical 
quantities, an exact analysis of a 
zero-dimensional field theory could help to obtain
information about the large order terms behavior in the perturbative 
series of realistic models. In particular, we may have in mind the paper
of Bender and Wu \cite{Wu}, who have obtained the precise asymptotic behavior
for $n\rightarrow \infty$ of the $n$-th order Rayleigh-Schr\"odinger 
coefficient in the series for the energy levels of the anharmonic oscillator. 
A precise knowledge of these coefficients for arbitrary $n$ is still 
missing, only bounds for their absolute values are available for 
finite generic values of $n$, even large.

Zinn-Justin \cite{zz} studying perturbation 
around instantons has used explicitly  the zero dimensional  
$\lambda\varphi^{4}$ model in order to introduce some basic ideas to 
perform a detailed analysis based on numerical 
simulations, of the 
large order behavior of the perturbative expansion of various 
models.
As noticed by Parisi \cite{Parisi1} and 
also by Khuri \cite{kuri} the nature of the large order 
estimates is strongly
dependent on the analytic structure of the (presumably) 
summed perturbative series $F(\lambda)=\sum_{n}f_{n}\lambda^{n}$
as a function of the coupling constant $\lambda$. 
A pioneering work on the subject
was done by Dyson \cite{Dyson}, who remarked that for  negative 
$\lambda=e^{2}$ in $(QED)_{4}$ the vacuum
is metastable with a meanlife $\sim e^{-\frac{1}{|\lambda|}}$, and that a 
cut along the negative real $\lambda$-axis  
is present. 
Dyson emphasizes that since any 
physical quantity evaluated by perturbative methods is not analytic at 
vanishing 
value of the coupling constant, the asymptotic expansion 
is not sufficient to determine the quantity uniquely.   
Parisi, \cite{Parisi1} using a functional representation of $F(\lambda)$
for $Re(\lambda)<0$,
still makes the important observation that a detailed knowledge of the 
imaginary
part at negative values of $\lambda$ would be necessary to improve 
Dyson's work. In other words,
to obtain more detailed estimates for the coefficients in 
the perturbative series, a better control of the imaginary part 
of $F(\lambda)$ for negative $Re(\lambda)$ is required.
It is worthwhile to remark that non-hermitian or unbounded  
Hamiltonians, in particular 
$(i\lambda\varphi^{3})$ and $(-\lambda\varphi^{4})$ models have been recently 
investigated by Bender et al. 
\cite{milton}. Arguing from the fact that  the model $(-\lambda\varphi^{4})$ is asymptotically free, these authors 
suggest that this theory should be useful for describing the 
Higgs boson. Historically, this idea of investigating the negative coupling 
scalar model in 
view of implementing asymptotic freedom is in fact present 
in the literature since the 70's \cite{brandt}.

In another branch or theoretical physics,
zero-dimensional field models can have a direct 
interest to the study of 
disordered systems, in particular to systems presenting frustration, which is associated to negative couplings in field theory language. These systems have been studied, both from a diagrammatic lattice 
viewpoint (quenched random graphs) by for instance Bachas et al. \cite{Claude}
and Baillie et al. \cite{Johnston}, or on 
more rigorous mathematical grounds by Derrida \cite{Derrida} 
and Aizeman et al. \cite {Ruelle}.

Perhaps an exact solution of the zero-dimensional 
$\lambda\varphi^{4}$ model could 
throw some light on the above described  situations.
In any case, it is clear that a main step to these kind of studies should be to understand how correlation functions behave for {\it complex} coupling constant, in particular for complex coupling constants having a {\it negative} real part. In this note we intend to go in this sense, by reducing an interacting system to its simplest possible form, the zero-dimensional $\lambda\varphi^{4}$ model. 
As a counterpart, an exact treatment is possible.

This paper is organized as 
follows. In section {\bf 2} the basic features of zero-dimensional field theory are 
reviwed. In section {\bf 3} we exhibit a non-perturbative (exact form) for the 
zero dimensional partition function and the analytic continuation in 
the coupling constant to the whole complex plane is performed. Also the exact form of
the partition function is compared to the expression obtained from Borel summation 
of the perturbative series. In this paper we use the standard convention $\hbar=k_{B}=1$.
Concluding remarks are in section {\bf 4}.

\section {Zero-dimensional field theory}

Let $P(\varphi)\geq 0$ be a probability distribution over a random variable. The
moments $<\varphi^{n}>$ of the probability distribution are obtained from the 
generating function,
\begin{equation}
Z(J)=\int d\varphi P(\varphi)e^{J\varphi},
\label{funcao geratriz 1}
\end{equation}
by successive derivatives,
\begin{equation}
<\varphi^{n}>=\frac{\int d\varphi\,\varphi^{n}P(\varphi)}{\int d\varphi P(\varphi)}=
\frac{1}{Z(0)}\left[\frac{\partial^{n}Z(J)}{\partial J^{n}}\right]_{J=0}
\label{momentos}
\end{equation}

Suppose that $P(\varphi)$ in Eq.(\ref{funcao geratriz 1}) has
the general form 
\begin{equation}
P(\varphi)=e^{-\frac{1}{2}\varphi A^{-1}\varphi+ f(\varphi,\lambda)},  \label{probabilidade nao Gaussiana}
\end{equation}
$f(\varphi,\lambda)$ being a regular function depending 
 on some parameter $\lambda$ (coupling constant). Then using the identity 
$f(\frac{\partial}{\partial J})e^{J\varphi}= f(\varphi)e^{J\varphi}$ the generating
function may be written in the form,
\begin{equation}
Z(J)=Z(0)e^{f(\frac{\partial}{\partial J},\lambda)}e^{\frac{1}{2}JAJ}=
Z(0)\sum_{n=0}^{\infty}\frac{1}{n!}\left[f(\frac{\partial}{\partial J})\right]^{n}e^{\frac{1}{2}JAJ},
\label{perturbacao}
\end{equation} 
which generates the diagrammatic expansion.

In this note we consider the model with a quartic probability 
distribution in which the partition function is given by, 
\begin{equation}
Z(m^{2},g)=\int_{-\infty}^{\infty}\frac{d\varphi}{\sqrt{2\pi}}
e^{-\frac{m^{2}}{2}\varphi^{2}-\frac{g}{4!}\varphi^{4}}.
\label{probabilidade quartica 1}
\end{equation}
The even order moments of this probability distribution can be obtained by successive derivatives respect to $m^{2}$,
\begin{equation}
<\varphi^{2n}>=\frac{\partial^{n}}{\partial (m^{2})^{n}}\ln Z(m^{2},g).
\label{momentos pares}
\end{equation}

The partition function given by Eq.(\ref{probabilidade quartica 1})
has a contribution from the vacuum diagrams, and its  
 perturbative expansion  may be written in the form \cite{itzik},
\begin{equation}
Z(m^{2}=1,g)=\sum_{n=0}^{\infty}(-g)^{n}z_{n},
\label{coef}
\end{equation}
where the coefficients are given by
\begin{equation}
z_{n}=\frac{(4n-1)!!}{(4!)^{n}n!}.
\label{coeff}
\end{equation}

Since the coefficient $z_{n}$ increases as $n^{n}$, the 
series that defines the partition function is divergent.
In this perturbative context, many authors claim that the 
point $g=0$ is an essential singularity of $Z(m^{2}=1,g)$, with 
a cut on the negative real axis. In other words, the 
perturbative series may have zero radius of convergence. Nevertheless, 
resummation techniques can be used to deal with this non-convergent 
series. It is important to stress that in real models in field 
theory the same kind of problem appears when the perturbative 
series is asymptotic but divergent for finite values of the 
coupling constant. Actually, 't Hooft and Lautrup \cite{t} showed that in the
$\lambda\varphi^{4}$ model the Borel transform of the perturbative 
series has renormalons, which prevents  Borel 
summability. In this case an alternative method which takes into account 
the existence of renormalons was developed by Khuri \cite{kuri}.

In the zero dimensional model, although for negative $g$ 
the partition function is not defined, an analytic continuation 
from positive $g$ can be performed by considering the contribution from the 
saddle points, as was remarked by Langer \cite{lan}. In this paper, we adopt  
 an easier way to obtain information from the region $Re(g)<0$.
Although the partition function, when $Re(g)<0$ is divergent, 
we are able to recover this divergence as singularities of
a function defined on the complex coupling constant plane.
In  other words, we will obtain first an exact expression in terms 
of Bessel functions of the second kind for the 
partition function in the domain $Re(g)>0$, and after this step, we 
analytically extend this function to the complex plane (i.e. also 
in the region where the original partition function diverges, $Re(g)<0$).
\textit{We take this analytic continuation to the whole coupling constant 
complex plane as our definition of the partition function}.

\section {The analytic continuation of the partition function}

We have two different steps to accomplish: 
the first one 
is to find a representation in terms of special functions 
of Eq.(\ref{probabilidade quartica 1}),
and the second one is to perform the analytic 
extension in the $g$-variable 
to the region $Re(g)<0$.
We accomplish our first step by a simple inspection in Gradhstein and Ryzhik \cite{Gradhstein}.   
Integrals of the type in Eq.(\ref{probabilidade quartica 1}) can 
be expressed as
\begin{equation}
\int_{-\infty}^{\infty}dxe^{-2\nu x^{2}-\mu x^{4}}=\frac{1}{2}\sqrt{\frac{2\nu}{\mu}}
e^{\frac{\nu^{2}}{2\mu}}K_{\frac{1}{4}}(\frac{\nu^{2}}{2\mu}),
\label{Bessel}
\end{equation}
for $Re(\mu)>0$,
which means that the partition function given by
Eq. (\ref{probabilidade quartica 1}) may be exactly expressed in terms of the Bessel function of the second kind $K_{\frac{1}{4}}$ (Macdonald's function)
in the form,
\begin{equation}
Z(m^{2},g)=\sqrt{\frac{2}{\pi}}\sqrt{\frac{3m^{2}}{4g}}
e^{\frac{3m^{4}}{4g}}K_{\frac{1}{4}}(\frac{3m^{4}}{4g}),
\label{Z-Bessel 1}
\end{equation}
in the domain $Re(g)>0$. Defining a rescaling of the coupling constant,
$\lambda=\frac{4g}{3m^{4}}$, the partition function becomes
\begin{equation}
Z(m^{2},\lambda)=\sqrt{\frac{2}{\pi}}\frac{1}{\sqrt{m^{2}\lambda}}
e^{\frac{1}{\lambda}}K_{\frac{1}{4}}(\frac{1}{\lambda}),
\label{Z-Bessel 2}
\end{equation}
valid for $Re(\lambda)>0$. An inspection of  
Eq. (\ref{Z-Bessel 2}) seems to indicate in the   
analytic structure of $Z(m^{2},\lambda)$ the existence of an essential 
singularity at $\lambda=0$. Actually, as we will see later on,
this is only apparent, the only singularity present 
in the analytically continued partition
function is a branch cut for values of $\lambda$ lying along the negative real axis.
We will perform our second step, by analytically extending the partition function
$Z(m^{2},\lambda)$ to the region $Re(\lambda)<0$, i.e.
to the whole complex $\lambda$-plane.

This analytical continuation may be done by simply 
 starting from the following representation 
for the Bessel functions of the second kind \cite{Lebedev},
\begin{equation}
K_{\nu}(z)=\sqrt{\frac{\pi}{2z}}\frac{e^{-z}}{\Gamma(\nu +\frac{1}{2})}
\int_{0}^{\infty}ds\,e^{-s}s^{\nu -\frac{1}{2}}\left[1+\frac{s}{2z}\right]^{\nu -\frac{1}{2}},
\label{continuacao 1}
\end{equation}
valid for $|arg(z)|<\pi$, $Re(\nu)>-\frac{1}{2}$.
Replacing the above equation in Eq.(\ref{Z-Bessel 2}) we obtain an analytic continuation for $Z(m^{2},\lambda)$ in the whole complex $\lambda$-plane
except for a cut along the negative real $\lambda$-axis,
\begin{equation}
Z(m^{2},\lambda)=\frac{1}{m\Gamma(\frac{3}{4})}\int_{0}^
{\infty}ds\,e^{-s}s^{-\frac{1}{4}}\left[1+
\frac{s\lambda}{2}\right]^{-\frac{1}{4}},
\label{Z-Bessel 3}
\end{equation}
for $|arg(\lambda)|<\pi$.

We have thus as a starting point a formula  
for the partition function defined for $Re(\lambda)>0$. 
It happens that this function has a representation defined
as an analytic function 
on the domain $|arg\, \lambda|<\pi$. Hence, we have an 
analytic extension of the partition function 
for the whole complex plane of the coupling constant $\lambda$,
except for $|arg\,\lambda|=\pi$. In other words, we have
an exact expression for the partition function valid in the
whole coupling constant complex plane except for a branch cut
on the real negative axis.
Actually, we may obtain in a closed form an expression
for the partition function. From the representation of
Macdonald's function in terms of the confluent hypergeometric 
function,
\begin{equation}
K_{\nu}(z)=\sqrt(\pi)(2z)^{\nu}e^{-z}\Psi(\nu+{\frac{1}{2}},
2\nu + 1,2z),
\label{confluent_1}
\end{equation}
we obtain replacing the above representation in Eq. (\ref{Z-Bessel 2})
the simple expression,
\begin{equation}
Z(m^2,\lambda)={\frac{1}{m}}({\frac{2}{\lambda}})^{\frac{3}{4}}
\Psi({\frac{3}{4}},{\frac{3}{2}},{\frac{2}{\lambda}}).
\label{confluent_2}
\end{equation}
We remark that the $\Psi(a,c,z)$ is a many-valued function of $z$,
and we shall consider in the above equation
its principal branch in the plane cut along
the negative real axis. The analytic continuation of 
$K_{\frac{1}{4}}(\frac{1}{\lambda})$  
corresponds to the definition of $Z(m^{2},\lambda)$ on
the whole complex $\lambda$-plane except for a branch cut
for $|arg\,\lambda|=\pi$.

The plots of the real and imaginary parts of the
analytically continued partition
function given by Eq. (\ref{Z-Bessel 3}) are in (fig.1)
and (fig.2).
We see from these figures that the real part of the partition function is
perfectly regular for any complex values of $\lambda$. 
The branch cut for $\lambda$ on the negative real axis appears only in the imaginary
part of the partition function. It does not appear in those graphics
an essential singularity at $\lambda=0$, as claimed by many 
authors. Indeed, using the expansion for the Bessel 
function of the second kind
for small $\lambda$, we get from Eqs. (\ref{Z-Bessel 3}), (\ref{continuacao 1}) and  
(\ref{Z-Bessel 2}) a Taylor series for the partition function, 
\begin{equation}
Z(m^2,\lambda)|_{m^{2}=1}=1-{\frac {3}{32}}\,\lambda+{\frac {105}{2048}}\,{\lambda}^{2}-{\frac {
3465}{65536}}\,{\lambda}^{3}+{\frac {675675}{8388608}}\,{\lambda}^{4}-
{\frac {43648605}{268435456}}\,{\lambda}^{5}+O({\lambda}^{6})
\label{assintotica}
\end{equation}
valid for $|arg(\lambda)|<\pi$, which clearly shows the absence of an 
essential singularity of the partition function at $\lambda=0$.

It is interesting to compare our exact result in Eq.(\ref{confluent_2})
or Eq.(\ref{Z-Bessel 3}) 
with the partition function obtained from perturbative methods as 
in Eqs.(\ref{coef}) and (\ref{coeff}). Since, as argued in \cite{itzik} 
 the series in Eq.(\ref{coef}) 
is asymptotic, we define its Borel transform as 
\begin{equation}
B(b)=\sum_{n=0}^{\infty}\frac{z_{n}}{n!}(-b)^{n},
\label{bo}
\end{equation}
and replacing $z_{n}$ from Eq.(\ref{coeff}) in the above 
expression we can show that the Borel transformed series $B(b)$ is 
convergent, given by an hypergeometric function, 
\begin{equation}
B(b)=F(\frac{1}{4},\frac{3}{4};1;-\frac{2b}{3}).
\label{bor}
\end{equation}

Then, from the Watson-Nevanlinna-Sokal theorem \cite{Ne}, 
the divergent series 
in Eq.(\ref{coef}) is Borel summable and, remembering 
that $\lambda=\frac{4g}{3}$ and using Eq. (\ref{bor}),
its Borel sum is given by
\begin{equation}
Z_{pert}(m^{2}=1,\lambda)=\frac{4}{3\lambda}\int_{0}^{\infty}
db\, e^{-\frac{4b}{3\lambda}}B(b)=
\sqrt{\frac{2}{\pi}}\frac{1}{\sqrt{\lambda}}
e^{\frac{1}{\lambda}}K_{\frac{1}{4}}(\frac{1}{\lambda}).
\label{borr}
\end{equation}
The above representation for the partition function 
obtained using the Borel summation technique is 
valid for $Re(\lambda)>0$, $\lambda$ belonging 
to a disc $C_{R}=\{\lambda: Re(\lambda^{-1})>
\frac{1}{R}\}$, that is, the Borel summed expression for the partition 
function is restricted to positive values of the real part 
of the coupling constant. In this region the Borel summed
expression for the partition function coincides \textit{precisely}
with the exact partition function Eq. (\ref{Z-Bessel 2}) for $Re(\lambda)>0$.
For $Re(\lambda)<0$ the Borel summed partition function is no longer  valid.
The analytically continued exact expression given by Eq.(\ref{confluent_2}) 
or by E.(\ref{Z-Bessel 3}) should then be used.

The even order moments $\varphi^{2n}$ for all complex values of $\lambda$
except for $\lambda$ lying on the real negative axis 
can be obtained exactly from Eq.(\ref{confluent_2}) or from Eq.(\ref{Z-Bessel 3}) 
(remembering that
$\lambda=\frac{4g}{3m^{4}}$) by direct application of Eq.(\ref{momentos pares}). We note, in connection to the idea of perturbation around some kinds of  
non-Gaussian probability distributions (instantons)
in the paper by Zinn-Justin \cite{zz}, that we can
work out perturbation theory around the quartic 
probability distribution
Eq.(\ref{probabilidade quartica 1}) using its exact analytically extended
expression (\ref{confluent_2}) or (\ref{Z-Bessel 3}), 
not necessarily restricted 
to positive real
$\lambda$ values. Then a possible extension of this work is to introduce a $\sigma\varphi^6$
term in the partition function to discuss the tricritical singularity in this 
oversimplyfied model \cite{Gino}.

\section{Conclusions}

In quantum field theory it is well known that the separation of the 
Hamiltonian into the free and the interaction part leads to conceptual problems 
in  many models, since the perturbative expansion based on the 
free part is divergent. In these situations the interaction part should not be used 
as a small perturbation, because at the origin of the interaction parameter  
an essential singularity would be present. 
The main idea is to include the interaction part in the new 
definition of a unperturbated Hamiltonian. It is expected that if it is possible 
to implement such a program, it would be equivalent to the resummation
of the perturbative series taking into account non-perturbative 
effects. In this paper we have used this idea to solve the zero 
dimensional $\lambda\varphi^{4}$ model.
Using the principle of analytic continuation, 
we have obtained an exact expression for the 
partition function of the model defined on the complex coupling constant plane,
which presents no essential singularity at the origin. 
We compare the Borel summed form
$Z_{pert}(m^{2}=1,\lambda
)$ of the partition function and our 
exact expression and we 
have a perfect agreement between the two functions in their common domain 
of validity.

\section{Acknowledgments}

We would like to thank L.H. Ford for a critical reading of the manuscript and B.F.Svaiter and I.Shapiro for fruitful discussions.
We also thank the Brazilian national Research Council (CNPq/MCT) for financial 
support.

\newpage

\

\

Figure Captions

\

Figure 1: Plot of the real part of the partition function $Z(m^2,\lambda)$
in the complex coupling constant plane. We take $m^2=1$.

\

Figure 2: Plot of the imaginary part of the partition function $Z(m^2,\lambda)$
in the complex coupling constant plane. We take $m^2=1$. Note the branch
cut for $Re(\lambda)<0$.

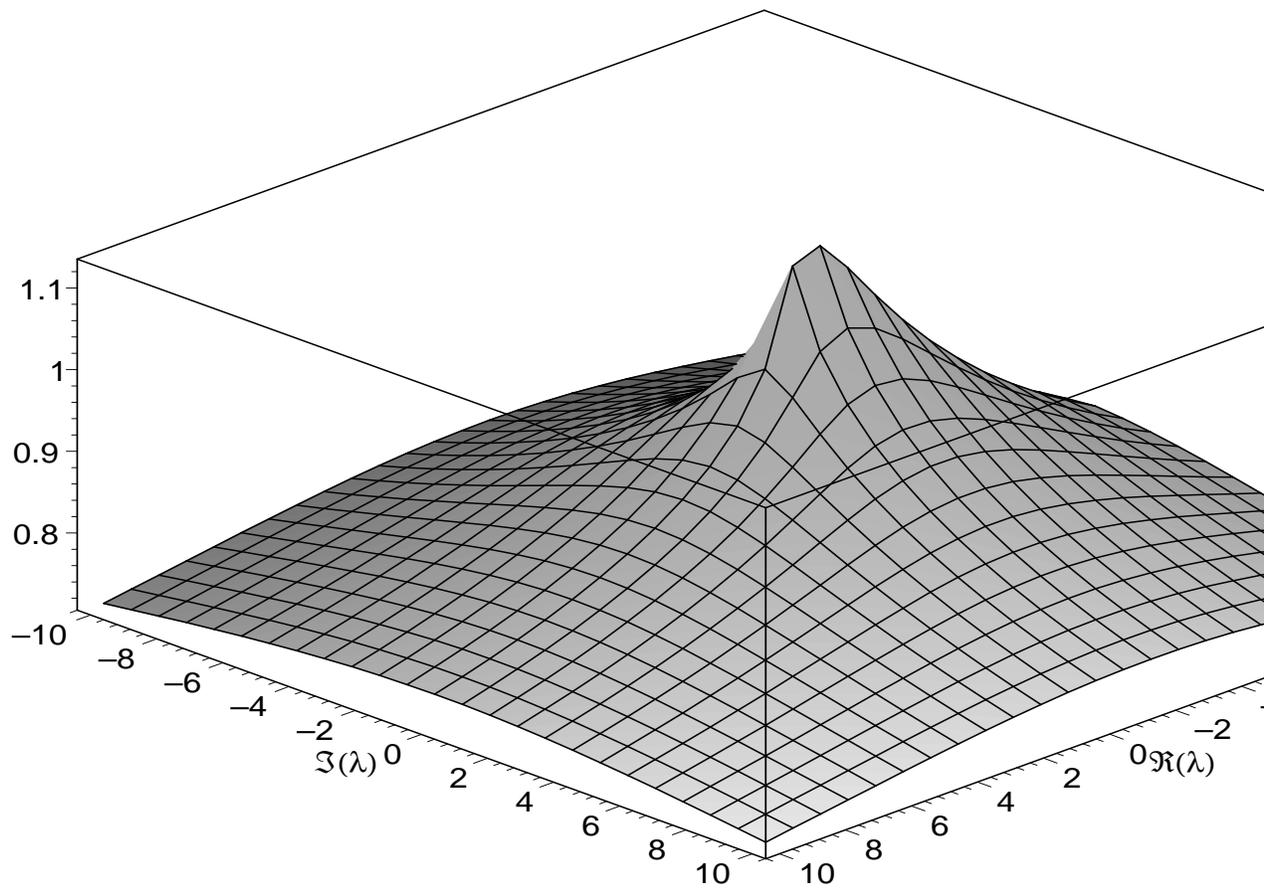
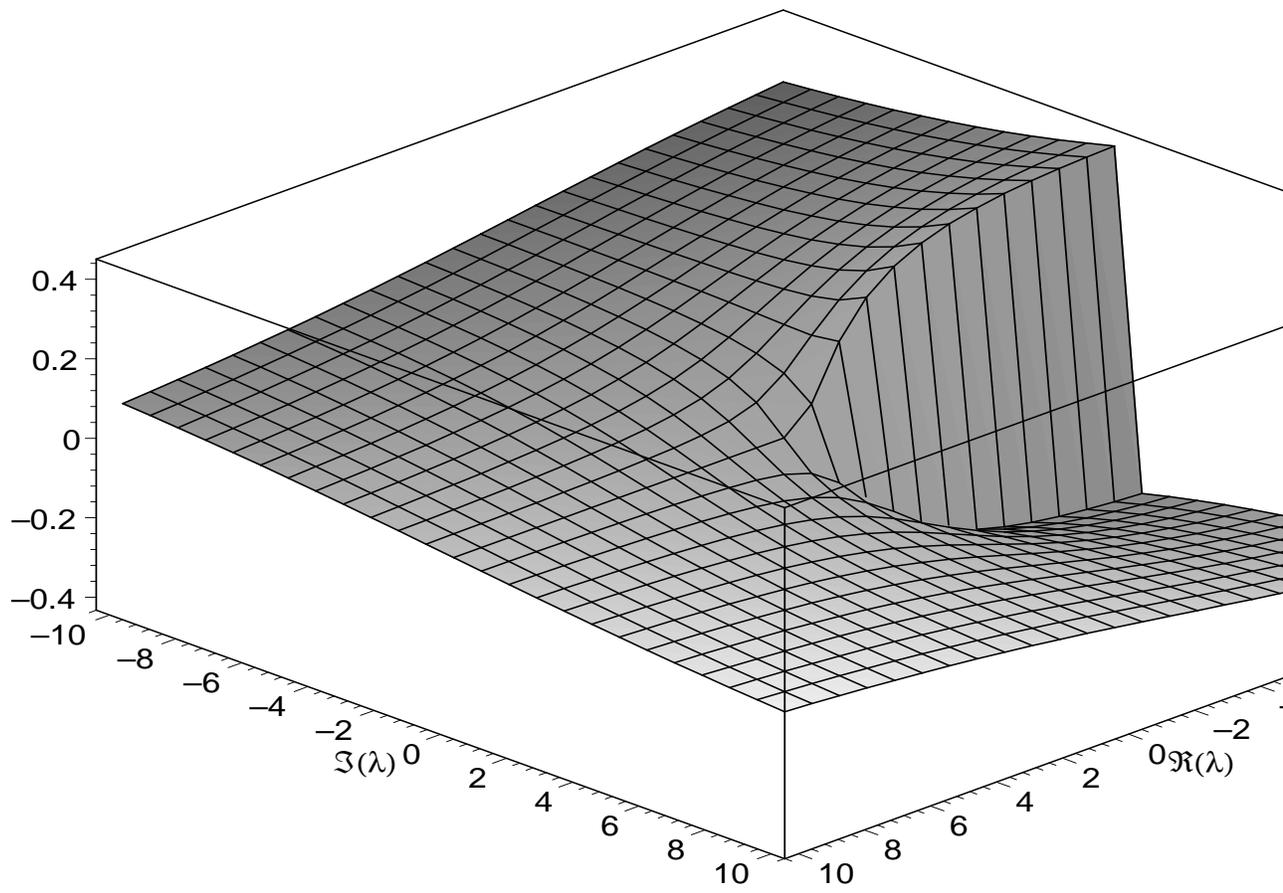
\end{document}